\documentclass[twocolumn,showpacs,preprintnumbers,amsmath,amssymb,aps]{revtex4}
\usepackage{graphicx}
\usepackage{dcolumn}
\usepackage{bm}
\usepackage[colorlinks=true]{hyperref}
\usepackage{epstopdf}
\usepackage[T1]{fontenc} 
\usepackage{lmodern}
\usepackage{braket}
\usepackage{color}

\begin{document}

\title{Dynamic spin polarization in organic semiconductors with intermolecular exchange interaction }

\author{A.~V.~Shumilin}

 \affiliation{
 Ioffe Institute, Russian Academy of Sciences, 194021 St.Petersburg, Russia
}

\begin{abstract} It is shown that in organic semiconductors where organic magnetoresistance (OMAR) is observed, the exchange interaction
between electrons and holes localized at different molecules leads to dynamic spin polarization in the direction
of the applied magnetic field. The polarization appears even at room temperature due to the non-equilibrium conditions.
The strong spin polarization requires exchange energy to be comparable with Zeeman energy in the external field and be larger or comparable with the energy of hyperfine interaction of electron and nuclear spins. The exchange interaction also modifies the lineshape of OMAR.
\end{abstract}

\maketitle
\section{Introduction}
\label{sec:intro}

Organic semiconductors represent a novel class of materials that attracts significant interest
nowadays. They are widely applied as light emitting diods~\cite{OLED,SmallMolOLED}.
The other possible applications are organic solar cells ~\cite{SolCells,Zhou2019,SmallMolOSC2020} and organic  transistors \cite{OFET1,OFET2}.
These semiconductors are promising materials for spintronics due to the long spin relaxation times and spin diffusion
lengths that can reach  dozens of nanometers \cite{Vard2004,SpinTimes,Drew2009}. In addition, the spin transport in organic semiconductors
is related to several intriguing phenomena that are not always well-understood.

The organic spin-valves are
unexpectedly easy to produce \cite{Dediu2009}. The conductivity of organic semiconductors is usually much
smaller than that of magnetic contacts that should exclude spin-valve magnetoresistance due to the spin injection \cite{Schmidt2000}. The thickness of the
devices often exceeds $100\,{\rm nm}$ and does not allow the tunneling through the organic layer \cite{Memristor}.
However, the
spin-valve magnetoresistance of the order of $10\%$ is measured in numerous experiments. Although an explanation
related to exchange interaction between carriers localized on different molecules was provided by Z.G. Yu \cite{Yu}, the reason of strong spin-valve magnetoresistance in
organic devices is still under discussion. While some groups report spin injection from magnetic contacts to
organic semiconductors, other groups consider this injection to be impossible \cite{Schmidt2013}. In this situation
the non-transport detection
of spin polarization in organics can be important. Such a detection was made with muon spin rotation and showed  the
existence of spin polarization in working organic spin-valve device \cite{Drew2009}.

Another interesting property of spin transport in organic semiconductors is the so-called organic
magnetoresistance (OMAR) \cite{kalin,OMAR0}. It is the strong
magnetoresistance observed in magnetic fields $\sim 10 - 100\,$mT both at low and room temperature.
In contrast with the strong organic spin-valve effect this phenomenon is generally understood.
Several mechanisms of OMAR were proposed \cite{prigodin, bobbert,OMAR-rev}. The magnetoresistance appears
because out of the equilibrium  the interaction between electrons and holes leads to correlations of electron and hole spins. These correlations can affect transport in organic semiconductors. Different mechanisms of OMAR are
related to different non-equilibrium processes including
exciton generation \cite{prigodin} and electric current combined with possibility of double occupation
of molecular orbitals \cite{bobbert}. The applied magnetic field suppresses the relaxation of spin correlations
that is caused by the hyperfine interaction of electron and nuclear spins \cite{YuHF}. It modifies the statistics of
spin correlations and leads to magnetoresistance.

The general understanding of OMAR requires only the interplay of non-equilibrium carrier statistics
and the hyperfine interaction of electron and nuclear spins. However, in some cases the exchange interaction between electrons
localized on neighbor molecules is invoked to describe the properties
of OMAR in particular materials \cite{IsotopeNat, IsotopePRB}.

In this paper it is shown that the interplay of the exchange interaction, hyperfine interaction  of electron and nuclear spins
and non-equilibrium phenomena in organic semiconductors leads to the polarization of electron spins. The polarization
occurs when Zeeman energy is comparable to the energies of hyperfine and exchange interactions. The temperature is
considered to be much larger than all these energies.
To the best of author's knowledge this phenomenon was never discussed in organic semiconductors. However the
similar effect was recently  observed in inorganic semiconductor quantum dots  \cite{Smirnov125,ourDP}
where the spin polarization can reach dozens of percents, and theoretically predicted in transition-metal dichalcogenides bilayers \cite{moire}. The effect was called the dynamic spin polarization
in contrast to the thermal spin polarization that requires Zeeman energy to be larger or comparable with temperature.

The paper is organized as follows. In Sec.~\ref{sec:model} the model of organic semiconductor is introduced. This model
includes the mechanisms of OMAR that are also responsible for the dynamic spin polarization when the exchange interaction
 is taken into account. In Sec.~\ref{sec:spinkin} the master equations
are derived that describe the spin dynamics due to hopping, hyperfine and exchange interaction. In Sec.~\ref{sec:OMAR}
the effect of the exchange interaction on conductivity and exciton generation rate is described. In Sec.~\ref{sec:pol} the dynamic spin polarization is obtained by the numeric solution of equations derived before. In Sec.~\ref{sec:res}
the specific ``resonance'' case is treated analytically. In Sec.~\ref{sec:dis} the general discussion and conclusion
of the results of this article are given.

\section{Model}
\label{sec:model}

Organic semiconductors are amorphous materials composed of single molecules or short polymers. The transport in
these materials is due to the hopping of electron and hole polarons over molecular orbitals \cite{Bassler,Baessler2012}.
Typically the organic semiconductors are strongly disordered due to the distribution of the energies of molecular orbitals with the width exceeding
$0.1\,{\rm eV}$ \cite{DOS2009,DOS2011} that is much larger than room temperature. Also the overlap integrals between neighbor
molecules differ in orders on magnitude \cite{BobbertAbin} enhancing the disorder in organic semiconductors.

The following model of organic semiconductor is adopted in this paper. Two molecular orbitals
in each molecule are considered: the highest occupied molecular orbital (HOMO) and
the lowest unoccupied molecular orbital (LUMO).
The charge transport is due to the
hoping of electron polarons over
LUMO, hopping of hole polarons over HOMO and in the
case of bipolar devices due to the generation of excitons from electron-hole pairs and their subsequent recombination.

The strong disorder localize the current in
rather sparse percolation cluster and the resistivity is controlled by rare bottlenecks in this cluster \cite{Shk,Cottaar}.
These bottlenecks are the pairs of molecular orbitals with relatively slow hopping rate between them.
In the case of bipolar devices the bottlenecks can
also be the pairs of HOMO and LUMO where electrons and holes recombine or form excitons.

The dynamic spin polarization is related to the same phenomena that lead to OMAR. The two most
known mechanisms of OMAR are the electron-hole (or exciton) mechanism and the bipolaron mechanism.
The electron-hole mechanism exists only in bipolar devices and is related to the different rates of singlet
and triplet exciton generation or to the different rates of recombination of electron and hole composing
singlet or triplet exciplet \cite{prigodin}. The bipolaron mechanism can also exist in unipolar devices but
requires the possibility of double occupation of molecular orbitals. It is assumed that double occupation
is possible only for electrons or holes in the spin-singlet state \cite{bobbert}.
To describe OMAR and dynamic polarization the theory of hopping transport that includes correlation of spins and occupation
numbers should be used. Such a theory developed in \cite{AVS2018, ShuBel1, ShuBel2, AVS2020} is applied in this paper.

\begin{figure}[t]
  \centering
  \includegraphics[width=0.5\textwidth]{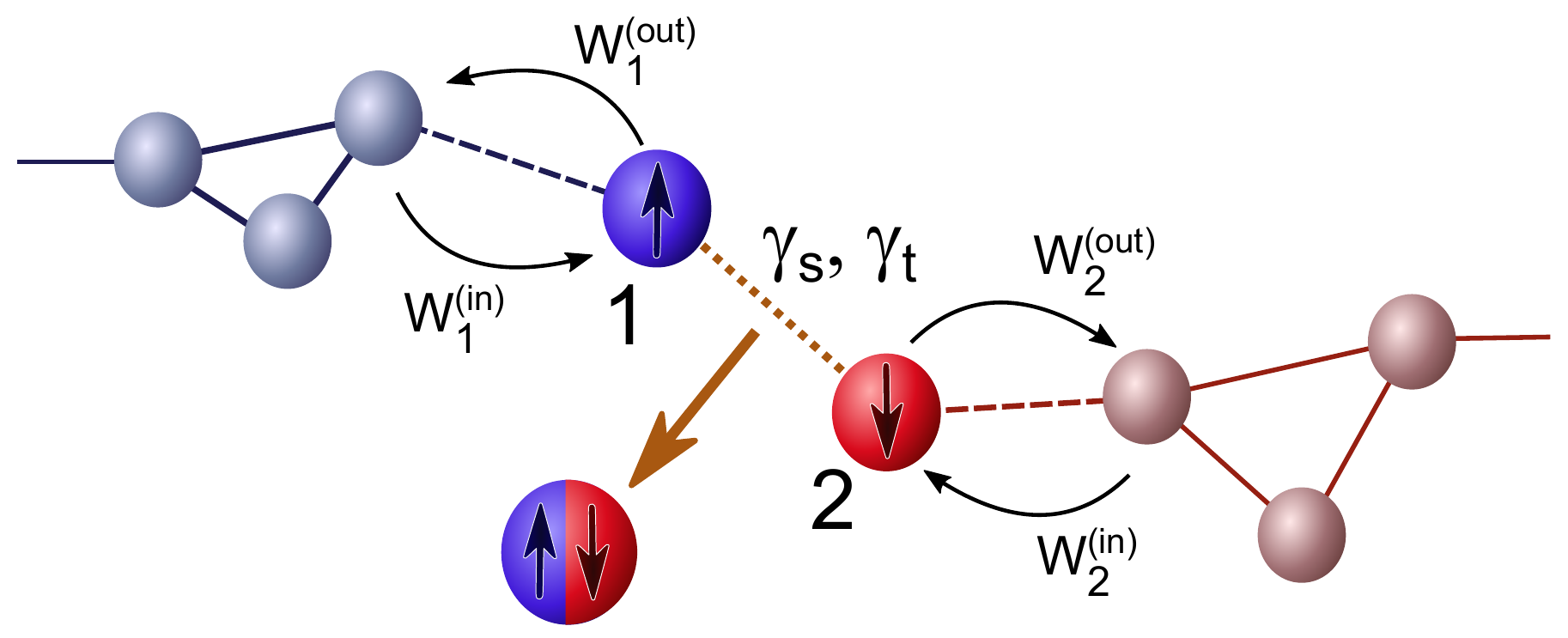}
  \caption{Electron-hole mechanism of OMAR. LUMO of molecules including the electron trapping site $1$ are
  shown with blue. HOMO including the hole trapping site $2$ are shown with red. Black arrows correspond to the rates of hopping to or from trapping sites. $\gamma_s$ and $\gamma_t$ are the rates of singlet and triplet exciton formation. The excitons never dissociate and recombine either radiatively or non-radiatively.  }
  \label{fig:eh}
\end{figure}

Both electron-hole and bipolaron mechanisms of OMAR are considered and the unified description for both
the mechanisms is given when possible.

In the case of electron-hole mechanism the bottlenecks that control
conductivity are considered to be the pairs of trapping sites for electron and hole (see Fig.~\ref{fig:eh}).
In such a pair the singlet exciton can be composed with the rate $\gamma_s$ and triplet one with the
rate $\gamma_t$.
After its generation the strong exchange interaction prevents exciton from changing its type.
The singlet excitons then recombine radiatively. Triplet excitons recombine either due to
phosphorescence \cite{phosphor1996,Baldo1998} or due to non-radiative processes. In the studied model the existing excitons do not affect the charge transport and formation of new excitons.  The current $J$ through the bottleneck
is proportional to the rate of exciton generation
\begin{equation}\label{J-eh}
J = e \gamma_s \frac{ \overline{n_1 n_2} - 4 \overline{s_1^{(\alpha)}s_2^{(\alpha)} }}{4} + e \gamma_t \frac{ 3\overline{n_1 n_2} + 4 \overline{s_1^{(\alpha)}s_2^{(\alpha)}} }{4}
\end{equation}
Here $\overline{n_1 n_2}$ is the probability of the joint occupation of LUMO site $1$ with an electron and HOMO site $2$
with a hole. $\overline{s_1^{(\alpha)}s_2^{(\beta)} }$ is the averaged product of spin polarization on site $1$
along Cartesian direction $\alpha$ and spin polarization on site $2$ along the direction $\beta$.
The sum over the repeating index $\alpha$ is assumed in Eq.~(\ref{J-eh}).

Without average spin polarizations $\overline{s_1^{(\alpha)}s_2^{(\beta)}}$ describe the correlations of spin directions. They can be expressed in terms of spin density matrix $\rho_s$ as follows:
 \begin{equation}\label{sasb-rho}
 \overline{s_1^{(\alpha)}s_2^{(\beta)}} = \frac{1}{4} {\rm Tr} \left[ \sigma_1^{(\alpha)}\sigma_2^{(\beta)} \rho_s \right]
 \end{equation}
Here $\sigma_1^{(\alpha)}$ is the Pauli matrix related to direction $\alpha$ and acting on the spin of trapping site $1$. $\sigma_2^{(\beta)}$ is the similar matrix for site $2$.

$\overline{s_1^{(\alpha)}s_2^{(\beta)}}$ are equal to zero in equilibrium because the temperature is large and $\rho_s$ is proportional to the identity matrix.  It will be shown in Sec.~\ref{sec:spinkin} that  $\overline{s_1^{(\alpha)}s_2^{(\beta)} }$ is proportional to $J$
  and the current can be expressed as follows
\begin{equation}\label{eh-J1}
J = e \gamma_{eh}(B) \overline{n_1 n_2}
\end{equation}
Here $\gamma_{eh}(B)$ is the effective exciton generation rate that depends on the applied magnetic field $B$.

The sites $1$ and $2$ are connected to other parts of the percolation cluster.
The electron can be trapped on molecule $1$ with the rate $W_{1}^{(in)}$ and be released with
rate $W_1^{(out)}$. $W_{2}^{(in)}$ and $W_{2}^{(out)}$ are similar rates for a hole to be trapped
on site $2$ and leave it respectively (see Fig.~\ref{fig:eh} for the directions of hops corresponding to these rates).
$W_{1,2}^{(in)}$ and $W_{1,2}^{(out)}$ are considered to be independent from magnetic field. It allows one to express
 magnetoresistance as a function of $\gamma_{eh}$. The corresponding derivations are provided in Appendix~\ref{appen-eh}.
When the magnetoresistance is relatively small, it can be expressed as follows
\begin{equation} \label{MR-gen}
\frac{R(B) - R(0)}{R(0)} = - \left\langle C_{eh} \frac{\gamma_{eh}(B) - \gamma_{eh}(0)}{\gamma_{eh}(0)} \right\rangle
\end{equation}
Here $R(B)$ is the sample resistance, $\langle...\rangle$ describes the averaging over bottlenecks where
 exciton generation occurs. $C_{eh}$ is the constant that is derived in the Appendix~\ref{appen-eh}.

\begin{figure}[t]
  \centering
  \includegraphics[width=0.4\textwidth]{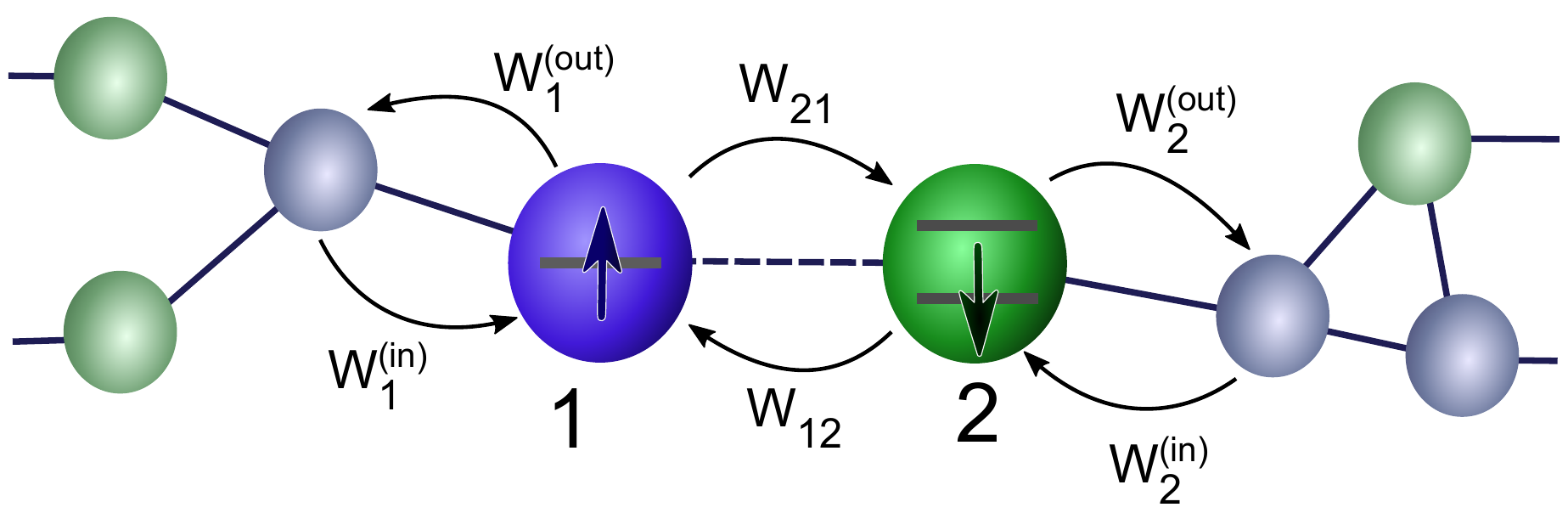}
  \caption{Bipolaron mechanism of OMAR. A-type LUMO are shown with blue color, B-type LUMO orbitals are shown with green color. Black arrows correspond to the rates of hopping to or from bottleneck.  }
  \label{fig:bp}
\end{figure}

The bipolaron mechanism of OMAR exists both in bipolar and unipolar devices. In this paper it
is discussed for the  unipolar devices with conductivity provided by electrons. It is assumed that
LUMO can be double occupied
by two electrons in spin-singlet state. The energy of double occupation is larger than the energy
of single occupation by the Hubbard energy $E_H \gg T$.  In this case all the molecular orbitals participating in
hopping transport can de divided into the two types.
The A-type orbitals can be unoccupied or single-occupied but are never double occupied due to the large Hubbard energy.
The B-type orbitals have very low energy of single occupation and therefore are always occupied by at least
one (resident) electron. Sometimes they are double occupied by electron pair in spin-singlet state. Note that both types
of orbitals are LUMO.

The bottlenecks in the unipolar transport are the
pairs of orbitals with the slowest hopping rates that are included into the percolation cluster.
Only the pairs of LUMO with different types are important for the bipolaron mechanism of OMAR.
Such a pair is shown in Fig.~\ref{fig:bp}. I call the LUMO in bottleneck the hopping sites $1$ and $2$ with analogy
to the trapping
sites in electron-hole mechanism. The site $1$ corresponds to $A$-type molecular orbital and site $2$ has type $B$
for definiteness.

The current in the bottleneck can be expressed as follows
\begin{multline}\label{bp-J0}
J = \frac{eW_{21}}{2} \left(  \overline{n}_1 - 4 \overline{s_A^{(\alpha)}s_B^{(\alpha)}}
- \overline{n_1n_2}  \right)\\
- 2eW_{12} \overline{n_2(1-n_1)}
\end{multline}
Here  $\overline{n_1 n_2}$ is the probability of joint occupation. $W_{12}$ and $W_{21}$ are the rates
of hops inside the bottleneck as shown in Fig.~\ref{fig:bp}.
Similarly to the electron-hole mechanism it will be shown that $\overline{s_1^{(\alpha)}s_2^{(\alpha)} }$ is
proportional to the current $J$. It allows to express $J$ as follows
\begin{equation}\label{bp-J1}
J = \kappa_{bp}(B) J_0,
\end{equation}
\begin{equation}
J_0 =  \frac{eW_{21}}{2}   \overline{ n_1 (1-n_2) } - 2eW_{12} \overline{n_2(1-n_1)}
\end{equation}
Here $\kappa_{bp}(B)$ is the probability of electron to be transferred through the critical pair.
It depends on the applied magnetic field leading to OMAR. When $|\kappa_{bp}(B) - \kappa_{bp}(0)| \ll \kappa_{bp}(0)$
the magnetoresistance can be expressed as follows
\begin{equation} \label{MR-gen-bp}
\frac{R(B) - R(0)}{R(0)} = -\left\langle \frac{\kappa_{bp}(B) - \kappa_{bp}(0)}{\kappa_{bp}(0)} \right\rangle
\end{equation}

\section{Spin dynamics}
\label{sec:spinkin}

In this section the dynamic of spin and spin correlations in the bottlenecks is described.
The temperature is considered to be large compared to Zeeman energy of electron spins,
energy of the hyperfine interaction and exchange energy, therefore in the equilibrium
there is no spin polarization or correlations of spin directions. The spin correlations appear
in the non equilibrium processes. For example if OMAR is controlled by electron-hole mechanism and the singlet
exciton formation is more probable than the formation of triplet exciton,
the triplet state of electron-hole pair in the bottleneck would be more probable than the singlet state.
I assume that when electrons and holes leave the bottleneck sites to the percolation cluster
they mix with other electrons and holes and the spin correlation is forgotten. It leads to an
effective spin relaxation. Finally the spin correlations have coherent dynamics in the bottleneck
due to hyperfine interaction with atomic nuclei and exchange interaction.

The kinetics of spin correlations can be expressed as follows:
\begin{multline}\label{scor-gen}
\frac{d \overline{s_1^{(\alpha)}s_2^{(\beta)}}}{dt} = \frac{i}{\hbar} \overline{[H, s_1^{(\alpha)}s_2^{(\beta)}]}
\\
+ G\delta_{\alpha \beta} - R_{\alpha\beta;\alpha'\beta'}\overline{s_1^{(\alpha')}s_2^{(\beta')}}
\end{multline}
Here the first term in r.h.s. describes the coherent spin dynamics due to
hyperfine interaction with atomic nuclei,
external magnetic field and the exchange interaction. $H$ is the Hamiltonian that includes all these energies.
$G\delta_{\alpha\beta}$ stands for the generation of spin correlations due to the
non-equilibrium processes. $R_{\alpha\beta;\alpha'\beta'}$ describes the relaxation
of spin correlations due to electron transfer between bottleneck and other parts of the percolation
cluster. $R_{\alpha\beta;\alpha'\beta'}$ also includes some contribution to relaxation of spin
correlations due to the incoherent processes inside the bottleneck.

In the bipolaron mechanism the generation of spin correlations is proportional to the current
\begin{equation}\label{bp-G}
G = \frac{J}{4e}
\end{equation}
and the relaxation is described with the expression
\begin{multline}\label{bp-R}
R_{\alpha\beta;\alpha'\beta'} = \left(W_{1}^{(out)} + W_2^{(in)}\right)\delta_{\alpha\alpha'}\delta_{\beta\beta'} \\ + \frac{W_{21}}{2} (\delta_{\alpha\alpha'}\delta_{\beta\beta'} - \delta_{\alpha\beta'}\delta_{\beta\alpha'}).
\end{multline}
This expression shows that spin correlation is forgotten when electron leaves $A$-type site $1$ to the
percolation cluster. The existence of the correlation assumes that $B$-type site $2$ is single
occupied. It cannot lose its last electron. However, the correlation is forgotten when the second electron comes to site $2$ from the percolation cluster while the site $1$ is occupied.
The hops from site $1$ to site $2$ are possible only in
the singlet state of the spins. Even without net current it leads to the relaxation of coherent combination
of singlet and triplet states \cite{AVS2020}. It is shown by the second term in r.h.s of Eq.~(\ref{bp-R}).

In the electron-hole mechanism of OMAR the spin correlations appear due to the different rates of singlet and triplet exciton formation
\begin{equation}\label{eh-G}
G = (\gamma_s - \gamma_t)\frac{\overline{n_1n_2} - 4\overline{s_1^{(\gamma)}s_2^{(\gamma)}}}{16}.
\end{equation}
Note that the probability of exciton decomposition is neglected, therefore the formation of exciton
 is possible only out of equilibrium. The relaxation of correlations for the electron hole
 mechanism of OMAR can be described as follows
\begin{multline} \label{eh-R}
R_{\alpha\beta;\alpha'\beta'} = (W_1^{(out)} + W_2^{(out)} + \gamma_t)\delta_{\alpha\alpha'}\delta_{\beta\beta'} +\\
 \gamma_s (\delta_{\alpha\alpha'}\delta_{\beta\beta'} - \delta_{\alpha\beta'}\delta_{\beta\alpha'}).
\end{multline}
This expression shows that the correlation is forgotten when electron leaves site 1 or hole leaves
site 2 to the percolation cluster. It is also transferred to the exciton in the process of triplet
exciton formation. The singlet exciton formation is similar to the hop from $A$-type site to $B$-type site
 in the bipolaron mechanism and leads to the relaxation of the coherent combinations
  of singlet and triplet states.

The hamiltonian $H$ can be expressed with the same equation for both the OMAR mechanisms.
\begin{equation}
H = H_B + E_{ex}{\bf s}_1{\bf s}_2
\end{equation}
Here $E_{ex}$ is the energy of exchange interaction of electron (or hole) spins on sites $1$ and $2$. $H_B$ describes the spin interaction with external magnetic field and atomic nuclei
\begin{equation}
H_B = \mu_b g \left({\bf B}+{\bf B}_{hf}^{(1)} \right) {\bf s}_1 + \mu_b g \left({\bf B}+{\bf B}_{hf}^{(2)} \right) {\bf s}_2
\end{equation}
Here ${\bf B}$ is the external magnetic field. ${\bf B}_{hf}^{(1)}$ and ${\bf B}_{hf}^{(2)}$  are the so-called hyperfine fields that
describe hyperfine interaction with atomic nuclei on sites $1$ and $2$ respectively.
It is presumed that on different sites the carrier spins interact with different nuclei,
therefore ${\bf B}_{hf}^{(1)}$ and ${\bf B}_{hf}^{(2)}$ are independent.
The distribution density  of hyperfine fields is
\begin{equation}
{\cal F}\left({\bf B}_{hf}^{(1,2)}\right) = \frac{1}{(\sqrt{2\pi}\Delta_{hf})^3}\exp \left( - \frac{\left|{\bf B}_{hf}^{(1,2)} \right|^2}{2\Delta_{hf}^2} \right)
\end{equation}
Here $\Delta_{hf} \sim 10\,{\rm mT}$ is the typical hyperfine field.
 The description of hyperfine interaction with static hyperfine field corresponds
 to the limit of many nuclear spins interacting with a single electron spin.

The interaction with external and hyperfine fields leads to precession of electron and hole spins
with frequencies ${\bm \Omega}_{1,2} = \mu_b g ({\bm B} + {\bm B}_{hf}^{(1,2)})/\hbar$ related to
sites $1$ and $2$
\begin{multline}
\frac{i}{\hbar}\overline{[H_B, s_1^{(\alpha)}s_2^{(\beta)} ]}  = \epsilon_{\alpha\gamma\alpha'} \Omega_1^{(\gamma)}\overline{s_1^{(\alpha')}s_2^{(\beta)}} \\ +
\epsilon_{\beta\gamma\beta'}
\Omega_2^{(\gamma)}\overline{s_1^{(\alpha)}s_2^{(\beta')}}
\end{multline}
The spin dynamics due to the exchange interaction can be described with the expression
\begin{multline}\label{s1s2-Jex}
\frac{i}{\hbar} E_{ex}  \overline{ [{\bf s}_1{\bf s}_2, s_1^{(\alpha)}s_2^{(\beta)}]}   = \\ \frac{1}{4} \frac{ E_{ex}}{\hbar} \epsilon_{\alpha\beta\gamma} \left( \overline{s_1^{(\gamma)} s_2^{(0)}} - \overline{s_1^{(0)} s_2^{(\gamma)}} \right)
\end{multline}
Here $\overline{s_1^{(0)}s_2^{(\gamma)}}$ describes the polarization of site $2$ in the direction $\gamma$ while the
site $1$ is single occupied.
\begin{equation}\label{s0sa-rho}
\overline{s_1^{(0)}s_2^{(\gamma)}} = \frac{1}{2} {\rm Tr} \left[ \sigma_1^{(0)}\sigma_2^{(\gamma)} \rho_s \right]
\end{equation}
which is similar to Eq.~(\ref{sasb-rho}). $\sigma^{(0)}_1$ is the unit matrix that acts on single occupied states of site $1$.
Therefore $s_1^{(0)}$ and $s_2^{(0)}$ are the operators of single occupation of sites $1$ and $2$ respectively.

For  the electron-hole mechanism or for  A-type sites in the bipolaron mechanism the ``single occupation'' and
``occupation'' are the same and  $s_{1,2}^{(0)} = n_{1,2}$.
In the considered model of bipolaron mechanism site $2$ is a $B$-type site and in this case $s_2^{(0)} = 1-n_2$ because it is single-occupied when the second electron is absent. In what follows the notation $s_{1,2}^{(0)}$ is used
when single-occupation is important for spin degrees of freedom, and $n_{1,2}$ is used for the description of current.

It is often assumed that statistics of spins is conserved when all the spins are reversed. If that would be the case
 $\overline{s_1^{(\gamma)} s_2^{(0)}}$ and $\overline{s_1^{(0)} s_2^{(\gamma)}}$ should be equal to zero.
 However, it will be shown that due to the exchange interaction even small external
 field $\sim 10-100\, {\rm mT}$ breaks the time reversal symmetry and leads to spin polarization in
 non-equilibrium conditions. To show it the kinetic equations for
 $\overline{s_1^{(\gamma)} s_2^{(0)}}$, $\overline{s_1^{(0)} s_2^{(\gamma)}}$ and the spin
 polarizations $\overline{s}_1^{(\alpha)}$, $\overline{s}_1^{(\alpha)}$ should be given.

I start from the expressions for spin polarizations in the electron-hole mechanism
\begin{multline}\label{s1}
\frac{d}{dt}\overline{s}_1^{(\alpha)} =  -\frac{E_{ex}}{\hbar} \epsilon_{\alpha\beta\gamma} \overline{s_1^{(\beta)} s_2^{(\gamma)}} + \epsilon_{\alpha\beta\gamma}\Omega_1^{(\beta)} \overline{s_1^{(\gamma)}} - W_{1}^{(out)}  \overline{s}_1^{(\alpha)} - \\
\gamma_t \overline{s_1^{(\alpha)} s_2^{(0)}}
+ \gamma_s \left(\overline{s_1^{(0)} s_2^{(\alpha)}} - \overline{s_1^{(\alpha)} s_2^{(0)}}  \right),
\end{multline}
\begin{multline}\label{s2}
\frac{d}{dt}\overline{s}_2^{(\alpha)} =  
\frac{E_{ex}}{\hbar} \epsilon_{\alpha\beta\gamma} \overline{s_1^{(\beta)} s_2^{(\gamma)}}
+ \epsilon_{\alpha\beta\gamma}\Omega_2^{(\beta)} \overline{s_2^{(\gamma)}} - W_{2}^{(out)} \overline{s}_2^{(\alpha)}- \\
\gamma_t \overline{s_1^{(0)} s_2^{(\alpha)}}
+ \gamma_s \left(\overline{s_1^{(\alpha)} s_2^{(0)}} - \overline{s_1^{(0)} s_2^{(\alpha)}}  \right).
\end{multline}

The first term in r.h.s. of Eqs.~(\ref{s1},\ref{s2}) describes the mutual precession of spins with
frequency $E_{ex}/\hbar$. The second term stands for the spin precession in local fields.
The third one shows that the spin polarization is lost when the electron or hole leaves the critical pair.  The electron and hole spins can also be lost due to the formation of spin-polarized triplet exciton with the rate $\gamma_t$. Note that the recombination of the spin on site $1$ requires the hole on site $2$ therefore the relaxation of spin on site $1$ is proportional to $\overline{s_1^{(\alpha)}s_2^{(0)}}$. The singlet exciton formation cannot relax total spin but leads to re-distribution of spin polarization between the sites with the rate $\gamma_s$.

When the bipolaron mechanism is considered, $1\rightarrow 2$ hop is an analogue of the singlet exciton formation
and $\gamma_s$ should be substituted with $W_{21}/2$. There is no analogue of triplet exciton formation in the
bipolaron mechanism and $\gamma_t$ should be substituted with zero in Eqs.~(\ref{s1},\ref{s2}).
Also $W_{2}^{(out)}$ should be substituted with $W_2^{(in)}$ in Eq.~(\ref{s2}) because $B$-type site cannot
lose it last electron but loses its spin polarization when it becomes double occupied.

\begin{figure}[t]
  \centering
  \includegraphics[width=0.4\textwidth]{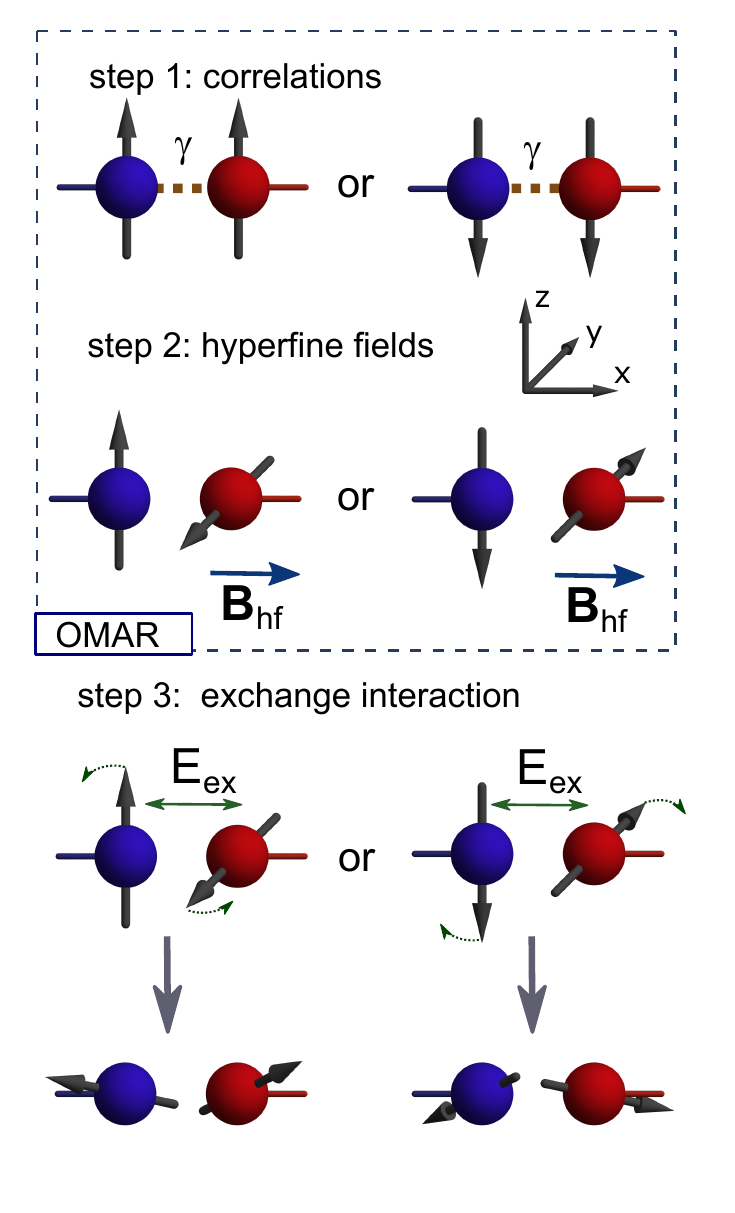}
  \caption{The three steps that lead to spin polarization of electrons and holes on the trapping sites. At the first step the spins become correlated due to the different rates of singlet and triplet exciton formation. At the second step the correlations are modified due to spin precession in effective on-site fields. It is shown with 90 degree rotation of the hole spin around x-axis. The modified correlations correspond to the coherent combination of singlet and triplet states. At the third step the spin polarization precess around the direction of the total spin. Initial  rotation directions and the spin directions after 90 degree rotation are shown. Spin polarization along $x$ axis of both electron and hole (averaged over left and right sides of the picture) appears.  }
  \label{fig:steps}
\end{figure}

Note that the exchange interaction leads to spin polarization only
when $\overline{s_1^{(\alpha)} s_2^{(\beta)}} - \overline{s_1^{(\beta)} s_2^{(\alpha)}} \ne 0$.
Such correlations correspond to the coherent combination of singlet and triplet states of the electron-hole pair.
Therefore the overall picture of dynamic spin polarizations can be described with the three
steps that are shown in Fig.~\ref{fig:steps}. At the first step the probabilities of singlet and triplet states
of polaron pair become non-equilibrium due to the different rates of singlet and triplet exciton formation or due
to the current and double occupation possibility. It means that spins of the carriers become correlated.
In Fig.~\ref{fig:steps} it is schematically shown with both the spins directed either up or down. It is important that electron  and hole spin polarizations averaged over left and right side of the figure are zero.

At the second step the spin precession with different frequencies ${\bm \Omega}_1$ and ${\bm \Omega_2}$
leads to a coherent combination of singlet and triplet states and to the correlations with
$\overline{s_1^{(\alpha)} s_2^{(\beta)}} - \overline{s_1^{(\beta)} s_2^{(\alpha)}} \ne 0$.
In Fig.~\ref{fig:steps} it is schematically shown with 90 degree rotation of hole spin on site 2 around $x$-axis.
These two steps are common for the theory of OMAR.

At the third step exchange interaction leads to mutual precession of
spins, or, which is the same, to precession of spins ${\bf s}_1$ and ${\bf s}_2$ around the direction of ${\bf s}_1 + {\bf s}_2$. In Fig.~\ref{fig:steps} the initial direction of this precession is shown together with the result of such a precession over angle $\pi/2$. The electron has negative polarization in $x$ direction both on the left and on the right side of the figure. It means that averaged electron spin polarization appears.

At this point the polarizations of electron and hole are opposite because the first terms in r.h.s. of Eq.~(\ref{s1}) and Eq.~(\ref{s2}) have equal absolute values and different signs. However, other terms in r.h.s. of Eqs.~(\ref{s1},\ref{s2}) are different and the  precession of spins with different
frequencies $\bm{\Omega}_{1}$ and $\bm{\Omega}_2$  and different rates of electron and hole spin
relaxation lead to non-zero averaged spin $\overline{\bf s}_1 + \overline{\bf s}_2$. It appears that usually
after the averaging over hyperfine fields the spin polarization on sites $1$ and $2$ have the same direction.

The kinetic equations for $\overline{s_1^{(\alpha)}s_2^{(0)}}$ and $\overline{s_1^{(0)}s_2^{(\alpha)}}$ are
similar to the equations for $\overline{s}_1^{(\alpha)}$ and $\overline{s}_1^{(\alpha)}$. However the terms
describing the transitions of electrons and holes between the bottleneck and other parts of the percolation cluster
are different
\begin{multline}\label{s1-0}
\frac{d}{dt}\overline{s_1^{(\alpha)} s_2^{(0)} } = 
- \frac{E_{ex}}{\hbar} \epsilon_{\alpha\beta\gamma} \overline{s_1^{(\beta)} s_2^{(\gamma)}} + \epsilon_{\alpha\beta\gamma}\Omega_1^{(\beta)} \overline{s_1^{(\gamma)} s_2^{(0)}}
\\ - (W_{1}^{(out)} + W_2^{(out)} + W_2^{(in)} + \gamma_t ) \overline{s_1^{(\alpha)} s_2^{(0)} }   \\
+ W_2^{(in)} \overline{s}_1^{(\alpha)}
+ \gamma_s \left(\overline{s_1^{(0)} s_2^{(\alpha)}} - \overline{s_1^{(\alpha)} s_2^{(0)}}  \right),
\end{multline}
\begin{multline}\label{s2-0}
\frac{d}{dt}\overline{s_1^{(0)}s_2^{(\alpha)}} = \frac{E_{ex}}{\hbar} \epsilon_{\alpha\beta\gamma} \overline{s_1^{(\beta)} s_2^{(\gamma)}}
+ \epsilon_{\alpha\beta\gamma}\Omega_2^{(\beta)} \overline{ s_1^{(0)}s_2^{(\gamma)}} - \\
 (W_{2}^{(out)}+ W_1^{(in)} + W_1^{(out)} + \gamma_t)\overline{s_1^{(0)}s_2^{(\alpha)}}  \\+ W_{1}^{(in)}\overline{s}_2^{(\alpha)}
+ \gamma_s \left(\overline{s_1^{(\alpha)} s_2^{(0)}} - \overline{s_1^{(0)} s_2^{(\alpha)}}  \right).
\end{multline}
To consider the bipolaron mechanism one should substitute $\gamma_s$ with $W_{21}/2$, $\gamma_t$ with zero and mutually exchange $W_{2}^{(in)}$ and $W_2^{(out)}$.

Eqs.~(\ref{scor-gen}-\ref{s2-0}) compose a system of 21 linear equations that should be solved
under stationary conditions together with Eqs.~(\ref{J-eh} - \ref{bp-J0}) that describe
electric current and exciton generation. However, in any case all the spin correlations and polarizations are proportional to $G$.
Therefore it is possible to express $\sum_\alpha \overline{s_1^{\alpha}s_2^{\alpha}} = {\cal T}_s G$ where ${\cal T}_s$
has the dimensionality of time. It allows one to use Eqs.~(\ref{eh-J1},\ref{bp-J1}).

In the electron-hole mechanism the rate $G$ of generation of spin correlation can be expressed in terms of
${\cal T}_s$ as follows
\begin{equation}
G = \frac{1}{16}\frac{(\gamma_s - \gamma_t) \overline{n_1 n_2}}{1 + {\cal T}_s (\gamma_s - \gamma_t)/4}
\end{equation}
Its substitution to Eq.~(\ref{J-eh}) leads to Eq.~(\ref{eh-J1}) with
\begin{equation}
\gamma_{eh} = \gamma_t + \frac{(\gamma_s - \gamma_t)}{4 + {\cal T}_s (\gamma_s - \gamma_t)}.
\end{equation}
In the bipolaron mechanism the similar arguments lead to the expression
\begin{equation}
\kappa_{bp} = \frac{1}{1 + W_{21} {\cal T}_s/2}.
\end{equation}

\section{Exciton formation rate and organic magnetoresistance}
\label{sec:OMAR}

Exchange interaction modifies the shape of OMAR and the dependence of
the exciton formation rate $\gamma_{eh}$ on applied magnetic field $B$. It can help to identify the
situations when dynamic spin polarization occurs in organic semiconductor.
Magnetoresistance and $\gamma_{eh}(B)$ dependence are related in the considered model in the electron-hole mechanism due to Eq.~(\ref{MR-gen}).
The shape of OMAR in the bipolaron mechanism has similar properties.
Therefore only the shape of $\gamma_{eh}(B)$
dependence is considered in this section for definiteness.

\begin{figure}[t]
  \centering
  \includegraphics[width=0.4\textwidth]{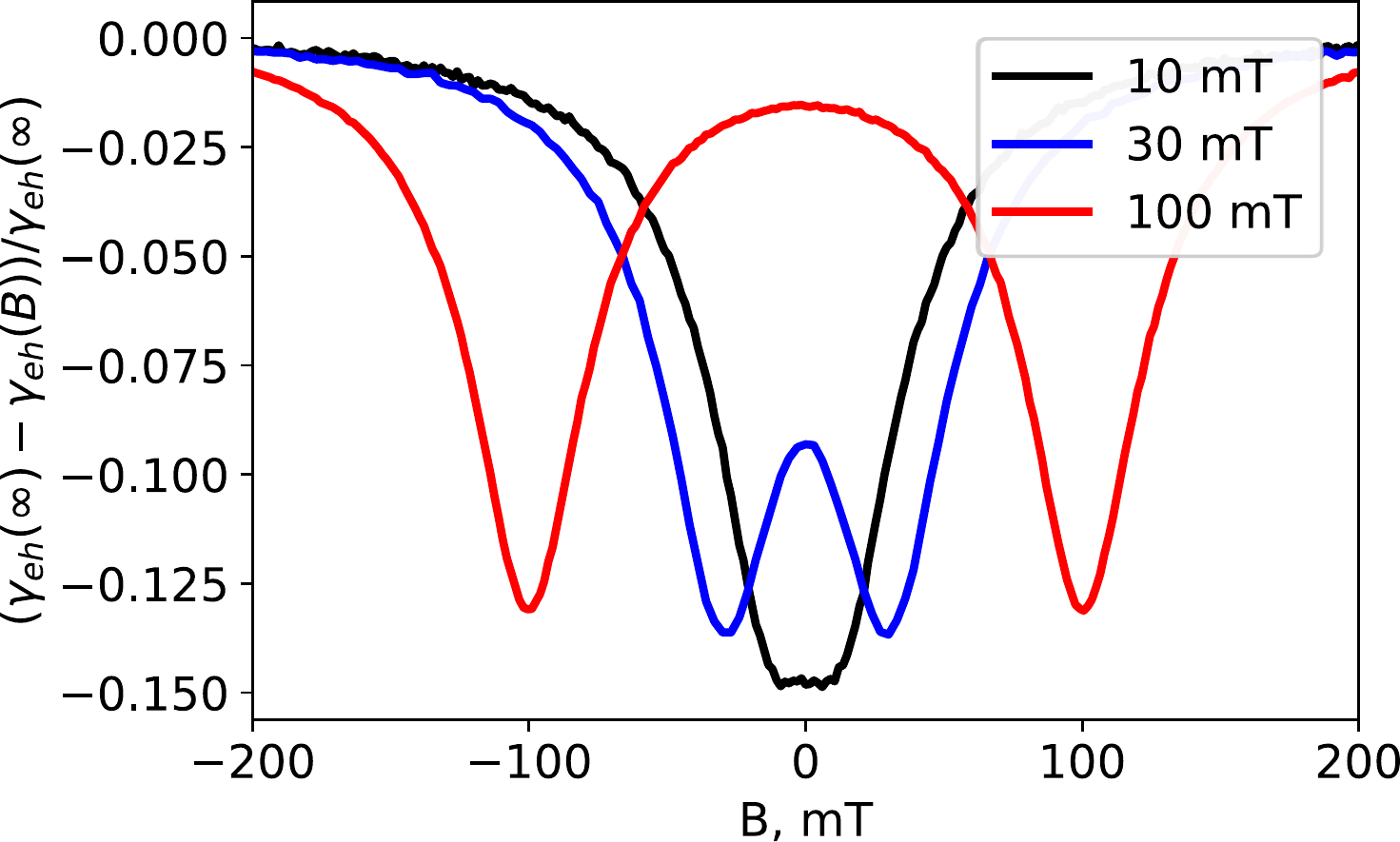}
  \caption{The shape of $\gamma_{eh}(B)$ dependence with different exchange fields $B_{ex}$ shown in legend. }
  \label{fig:OMAR1}
\end{figure}

The exciton formation rate  is calculated with  numeric solution of Eqs.~(\ref{scor-gen} - \ref{s2-0}).
The typical value of hyperfine field $\Delta_B = 10\,{\rm mT}$ is considered.
The exchange interaction significantly modify OMAR when the ``exchange field'' $B_{ex} = E_{ex}/g\mu_b$
is larger or comparable with the hyperfine field $B_{ex} \gtrsim \Delta_B$. Therefore the values of
$B_{ex}$ from $10$ to $100\,{\rm mT}$ are discussed in this section.

In Fig.~\ref{fig:OMAR1} the calculated dependence $\gamma_{eh}(B)$  is shown. It is compared to
 the value $\gamma_{eh}(\infty)$ in high magnetic fields where the exciton generation rate is saturated.
The  singlet and triplet exciton formation rates are considered to be  $\gamma_s = 20\, {\rm \mu s}^{-1}$ and
 $\gamma_t = 4\,  {\rm \mu s}^{-1}$ respectively. These rates are $\sim 10$ times slower than the spin precession
 in the hyperfine field. The transitions of electrons and holes between the sites $1$ and $2$ and
 the rest of percolation cluster are described with the rates $W_{1}^{(in)} = 30 \, {\rm \mu s}^{-1}$,
 $W_{2}^{(in)} = 100 \, {\rm \mu s}^{-1}$, $W_{1}^{(out)} = 5 \, {\rm \mu s}^{-1}$, $W_{2}^{(out)} = 1 \, {\rm \mu s}^{-1}$.
 The relatively small values $W_{1}^{(out)}$ and $W_2^{(out)}$ show that the sites $1$ and $2$ are effective traps
 for electron and hole respectively. These parameters were considered to be the same for all the bottlenecks that control the transport. The averaging was made over $10^4$ random values of hyperfine fields ${\bf B}_{hf}^{(1)}$ and
 ${\bf B}_{hf}^{(2)}$.

\begin{figure}[t]
  \centering
  \includegraphics[width=0.4\textwidth]{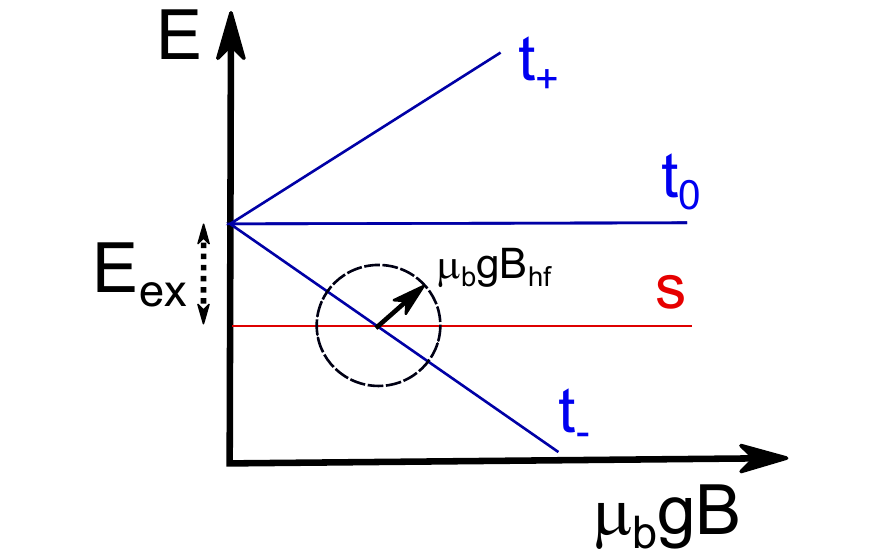}
  \caption{Energies of the triplet and singlet spin states in magnetic field. The circle shows where the
    hyperfine interaction can efficiently mix the states. }
  \label{fig:levels}
\end{figure}

The exchange interaction splits zero-field peak of $\gamma_{eh}(B)$ dependence. This splitting is small when ${B_{ex} \lesssim \Delta_B}$.
This results qualitatively agrees with the calculations made in Ref.~\cite{IsotopePRB} where it was compared with
magnetoresistance measured in $\rm Alq_3$.
When the exchange interaction is strong $B_{ex} \gtrsim 10\Delta_B$, $\gamma_{eh}(B)$ shape consists
of two peaks at $B = \pm B_{ex}$. It can be explained with the following model (Fig. \ref{fig:levels}).
 At zero magnetic field the exchange interaction prevents the hyperfine interaction  from mixing
 singlet and triplet states. In this case the hyperfine interaction does not affect the exciton formation and current.
 However the energy of singlet state $E_s$ does not depend of magnetic field while the energy
 of one of the triplet states $E_{t-}$ decreases with $B$. When $|B-B_{ex}| \lesssim \Delta_{hf}$,
 the singlet-triplet mixing due to hyperfine field becomes effective.
 It leads to the peak in $\gamma_{eh}(B)$ dependence.

\begin{figure}[t]
  \centering
  \includegraphics[width=0.4\textwidth]{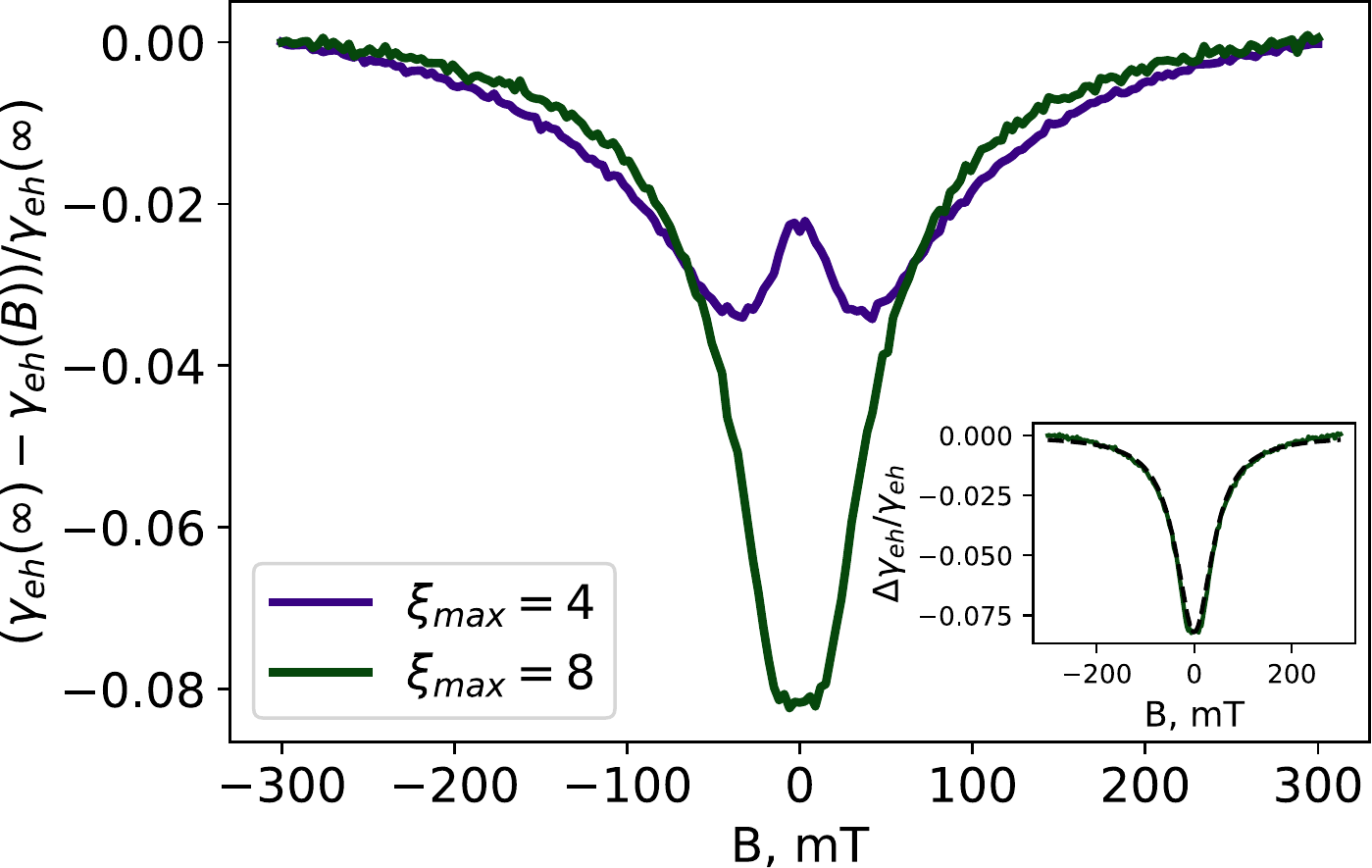}
  \caption{$\gamma_{eh}(B)$ dependence with the distribution of exchange fields described by Eqs.~(\ref{Ex-distr}) with different
  $\xi_{max}$ described in legend. The inset shows the comparison between the results for $\xi_{max} = 8$ (green curve) and Lorentz function
  $\gamma_{eh}(\infty)-\gamma_{eh}(B) \propto B^2/(B^2 + B_0^2)$ with $B_0 = 45\,{\rm mT}$ (black dashed curve). }
  \label{fig:OMAR-distr}
\end{figure}

To the best of the authors knowledge such a resonance OMAR shape was never observed in experiment. However,
it appears because the exchange interaction is considered to be the same for all the bottlenecks that control
exciton formation. In organic semiconductor the inter-molecular exchange interaction has strong dependence on
the overlap integrals that have the exponentially-broad distribution \cite{BobbertAbin}.
Therefore, one can expect the broad distribution of exchange energies. In real materials it should be determined with
numeric simulation. Here I consider only a simplified model that shows that exchange energy $E_{ex}$ in the bottleneck
can vary in order of magnitude
\begin{subequations}
\label{Ex-distr}
\begin{equation}
B_{ex} = B_{ex}^{(0)}\exp(-\xi)
\end{equation}
\begin{equation}
{\cal F}_{\xi}(\xi) = \frac{1}{\xi_{max}}, \quad \xi\in(0, \xi_{max})
\end{equation}
\end{subequations}
The exchange field is always smaller than $B_{ex}^{(0)}$. The value $\xi$ describes its suppression due to the small
overlap integral between sites $1$ and $2$. The distribution density ${\cal F}_{\xi}$ of the suppression factor is
flat between zero and $\xi_{max}$ that is the maximum suppression that still allows efficient
exciton generation. The shape of $\gamma_{eh}(B)$ dependence for $B_{ex}^{(0)} = 1T$ and $\xi_{max} = 4$ and $8$ is shown in Fig.~\ref{fig:OMAR-distr}.
The splitting of the peak is controlled by the smallest possible exchange energies. When $B_{ex}^{(0)}\exp(-\xi_{max})$
is large compared to $\Delta_B$ the splitting is clearly observable while for $B_{ex}^{(0)}\exp(-\xi_{max}) < \Delta_B$
it is suppressed and $\gamma_{eh}(B)$ dependence has Lorentz shape.

\section{dynamic spin polarization}
\label{sec:pol}

In the presence of exchange interaction the non-equilibrium phenomena that lead
to OMAR also yield the spin-polarizations of electrons and holes in the bottleneck. It can be
understood from non-zero values
of $d \overline{s}_1^{(\alpha)}/dt$ and $d \overline{s}_2^{(\alpha)}/dt$ in Eqs. (\ref{s1},\ref{s2}).  The relative polarizations on sites $1$ and $2$ are defined as $P_1$ and $P_2$ respectively.
\begin{equation}
P_1 = \frac{2\overline{s}_1^{(z)}}{\overline{s}_1^{(0)}}, \quad P_2 = \frac{2\overline{s}_2^{(z)}}{\overline{s}_2^{(0)}}.
\end{equation}
The polarizations are normalized to the single occupation probabilities of sites $1$ and $2$. The averaged polarization is always directed
along the axis of the applied magnetic field.

In the electron-hole mechanism of OMAR the produced triplet excitons are also spin-polarized and their polarization is
\begin{equation}
P_{ex} = \frac{\overline{s_1^{(z)}s_2^{(0)}} + \overline{ s_1^{(0)}s_2^{(z)}}}{\overline{n_1n_2}}.
\end{equation}

In both of the OMAR mechanisms spin current appears. It is different for the two parts of the percolation cluster connected to sites
$1$ and $2$ because the spin is not conserved in the bottleneck. I assume that the electrons that come from the percolation
cluster to the bottleneck are not spin-polarized. It leads to the following expressions for spin currents
$J_s^{(1)}$ and $J_s^{(2)}$
in the bipolaron mechanism
\begin{equation}
J_{s}^{(1)} = W_{1}^{(out)}\overline{s}_1^{(z)}, \quad J_{s}^{(2)} = -W_{2}^{(in)}\overline{s}_2^{(z)}.
\end{equation}
The similar expressions for the spin currents in the electron-hole mechanism  are
\begin{equation}
J_{s}^{(1)} = W_{1}^{(out)}\overline{s}_1^{(z)}, \quad J_{s}^{(2)} = -W_{2}^{(out)}\overline{s}_2^{(z)}.
\end{equation}

\begin{figure}[t]
  \centering
  \includegraphics[width=0.4\textwidth]{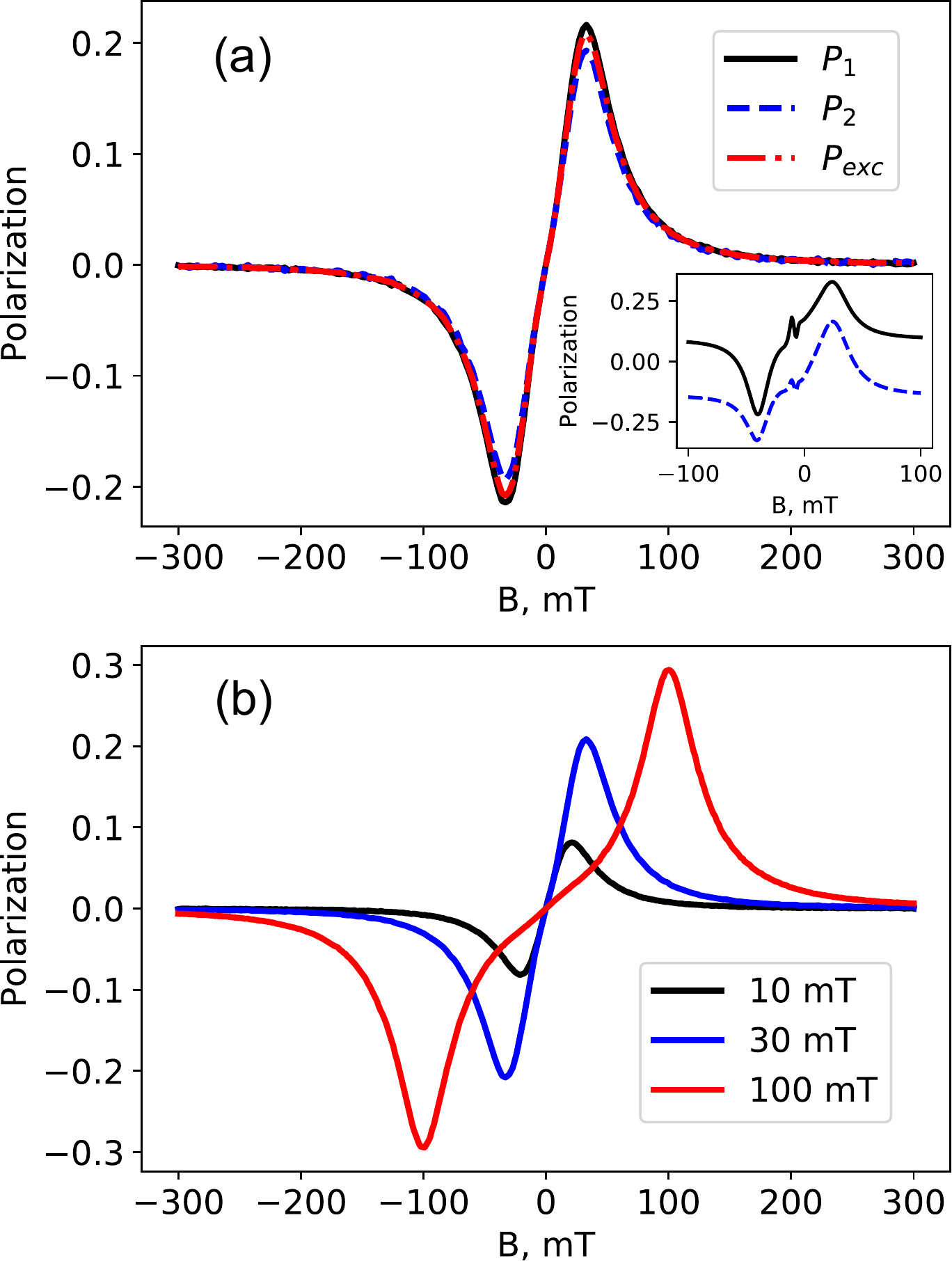}
  \caption{Dynamic spin polarization calculated numerically and averaged over $10^4$ hyperfine fields. (a) the spin polarization of electrons on site $1$, holes on site $2$ and
  triplet excitons for the exchange field $B_{ex} = 30\,{\rm mT}$. The inset shows the typical spin polarizations without averaging over hyperfine fields. (b) The triplet exciton spin polarization at different exchange fields shown in legend. }
  \label{fig:Pol1}
\end{figure}

Because the spin currents are directly related to spin polarizations, I discuss only $P_1$, $P_2$ and $P_{ex}$. In this section the spin polarizations are calculated numerically under stationary conditions for the systems described in Sec.~\ref{sec:OMAR}.
In Fig.~\ref{fig:Pol1}(a) magnetic field dependence of $P_1$, $P_2$ and $P_{ex}$ is shown
for the exchange field  $B_{ex} = 30\,{\rm mT}$ and the other conditions corresponding to Fig.~\ref{fig:OMAR1}.
The polarizations almost coincide in this situation. Some details about their coincidence are given in Sec.~\ref{sec:res}.  Note that although the signs of $d \overline{s}_1^{(\alpha)}/dt$ and $d \overline{s}_2^{(\alpha)}/dt$ in Eqs. (\ref{s1},\ref{s2}) are different when spin polarizations are zero, the resulting polarizations that take into account the spin transfer between molecules and (most important) the averaging over the hyperfine fields have the same sign. Inset in Fig.~\ref{fig:Pol1}(a) shows the typical spin polarizations $P_1$ and $P_2$ without the averaging over hyperfine fields. They have the same sign when $B$ is close to $\pm B_{ex}$ and different signs otherwise.

Fig.~\ref{fig:Pol1}(b) shows the
triplet exciton polarization at different exchange fields.
The shape of the dependence $P_{ex}(B)$ is almost independent of
$B_{ex}$ when $B_{ex}\le \Delta_{hf}$ while its amplitude grows with $B_{ex}$. For large exchange energy
the shape of the dependence contains the two  peaks at $B = \pm B_{ex}$ in agreement
with Fig.~\ref{fig:levels}.

\begin{figure}[t]
  \centering
  \includegraphics[width=0.4\textwidth]{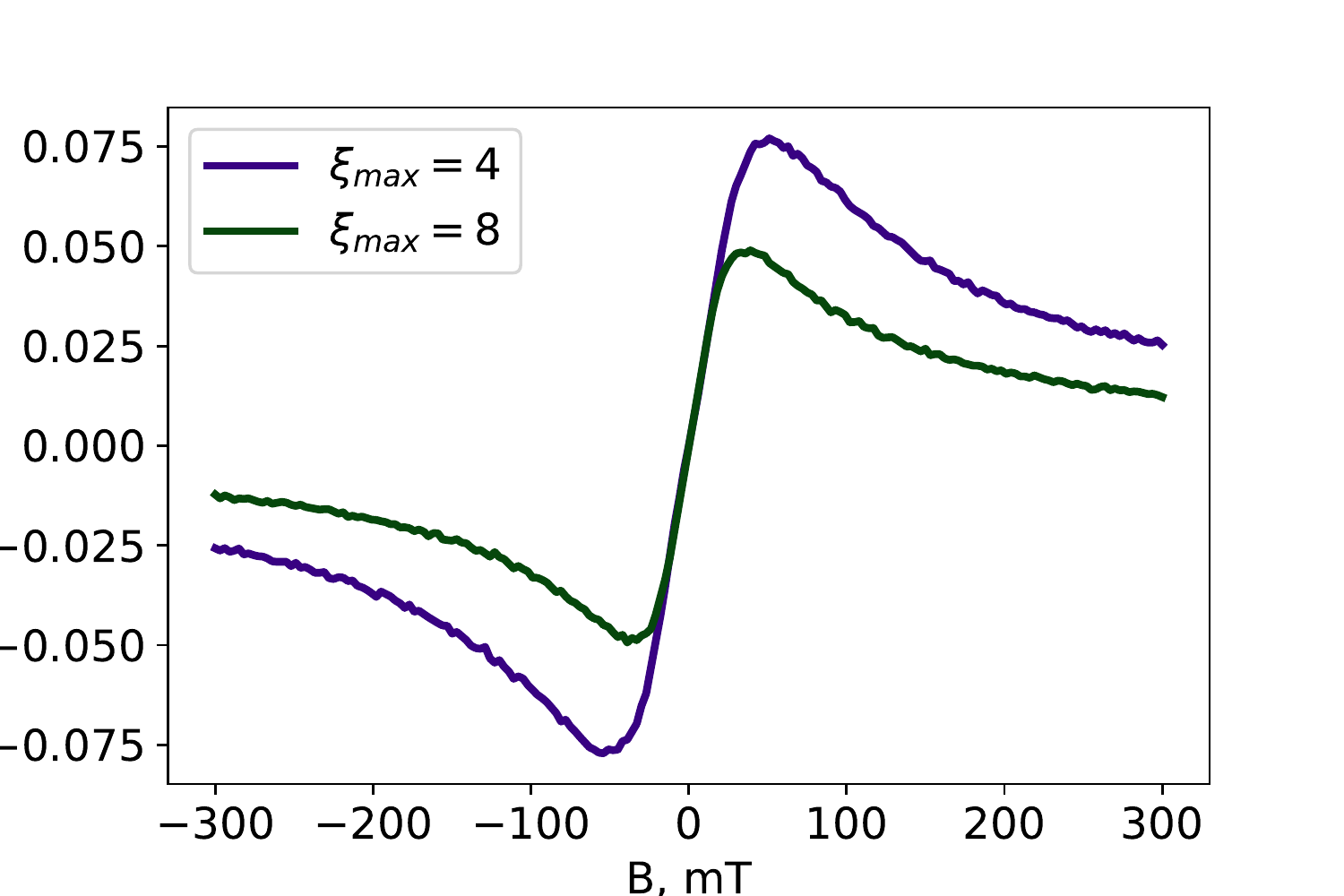}
  \caption{Dynamic polarization with the broad distribution of exchanged field described by Eq.~(\ref{Ex-distr}) with $\xi_{max}$ shown in legend.}
  \label{fig:Pol2}
\end{figure}

The results for the broad distribution of exchange energies described  with Eq.~(\ref{Ex-distr}) are shown
in Fig.~\ref{fig:Pol2}.
Similarly to $\gamma_{eh}(B)$ dependence shown in Fig.~\ref{fig:OMAR-distr} the peaks in the
 dependence of polarization on the applied field are smeared due to the distribution of the exchange energies.
However, even when the splitting of zero-field peak in $\gamma_{eh}(B)$ dependence is hardly
observable (as it is in the case of $\xi_{max} = 8$)
the spin polarization is not completely suppressed. The splitting of the zero-field peak depends on the minimal
possible exchange energy while the spin polarization in magnetic field $B$ depends on the
probability for the exchange field to be $B_{ex} \sim B$.

\section{Spin polarization in resonance}
\label{sec:res}

The results of numeric simulation provided in Sec.~\ref{sec:pol} show that spin polarization is the strongest in the
``resonance situation'' when $B = B_{ex} \gg B_{hf}$. This situation is studied in details in this section.
The electron-hole mechanism of OMAR is considered for definiteness.

To further simplify the theory I assume that singlet exciton formation rate is fast
$\gamma_s \gg W_1^{(in)} + W_1^{(out)} + W_2^{(in)} + W_2^{(out)}$ and triplet exciton formation rate is very slow
$\gamma_t \ll W_1^{(out)} + W_2^{(out)}$. The spin precession in hyperfine field is
fast compared to hops and exciton formation
rates $\mu_b g B_{hf} \gg \hbar \gamma_s$.

The singlet electron-hole pair forms exciton almost immediately after it appears.
The electron-hole pair in the state $t_-$ can easily change its state to singlet and also form a singlet exciton. However the pairs in states
$t_+$ and $t_0$ usually dissociate due to electron and hole leaving the sites $1$ and $2$. Sometimes, however, such a pair
forms a triplet exciton due to the finite rate $\gamma_t$. Even at this point it is clear that the triplet excitons are
strongly spin-polarized because no $t_-$ triplet excitons appear.

These assumptions allow to reduce the system (\ref{scor-gen} - \ref{s2-0}) for 21 ``spin variables'' and the joint equations
for ``charge'' variables $\overline{n}_1$, $\overline{n}_2$ and $\overline{n_1 n_2}$ to the system of six linear equations.
The new equations describe the system in terms of the following variables. $p_+$ and $p_0$ are the probabilities
for the bottleneck to be occupied by the electron-hole pair in $t_+$ and $t_0$ states respectively. The probabilities
of the bottleneck to be occupied by electrons and holes in singlet or $t_-$ states are neglected due to the
fast singlet exciton generation rate and effective coupling between singlet and $t_-$ state. $p_1$ and $p_2$ are
the probabilities for the occupation of sites $1$ and $2$ respectively while the second site is unoccupied. $p_{1s}$ is the
difference between the probabilities for site $1$ to be occupied by spin-up and spin down electron while the site $2$ is
empty. $p_{2s}$ is the similar quantity for the site $2$.

Polarizations $P_1$, $P_2$ and $P_{ex}$ are expressed in these notations as follows
\begin{subequations}\label{Pol-p}
\begin{equation}\label{Pol-p-ex}
P_{ex} = \frac{p_+}{p_+ + p_0},
\end{equation}
\begin{equation}\label{Pol-p-12}
P_1 = \frac{p_+ + p_{1s}}{p_+ + p_0 + p_1},
\quad P_2 = \frac{p_+ + p_{2s}}{p_+ + p_0 + p_2}.
\end{equation}
\end{subequations}
Eqs.~(\ref{Pol-p}) show that triplet exciton polarization is related only to the states when both the sites
of the bottleneck
are occupied, while the polarizations $P_1$ and $P_2$  are also affected by states when only one of the sites is occupied.

The system of equations for $p_\zeta$ where $\zeta = +,0,1,2,1s,2s$ follows from the stationary conditions $dp_\zeta/dt = 0$.
The stationary conditions for $p_+$ yields
\begin{multline}\label{6eq+}
\frac{p_1 + p_{1s}}{4}W_2^{(in)} + \frac{p_2 + p_{2s}}{4}
W_1^{(in)} \\ = (W_1^{(out)} + W_2^{(out)})p_+
\end{multline}
The electron-hole pair in $t_+$ state can appear when electron or hole is trapped  on the corresponding site of the
bottleneck while the second site is occupied. When the occupied site is not spin-polarized (for example when $p_1 = 1$, $p_{1s} =0$
and the hole becomes trapped on site $2$) any of the four spin states of electron-hole pair appears with equal probability.
When the site is fully spin-polarized ($p_1=p_{1s} = 1$) the probability of $t_+$ state after the hole trapping is $1/2$.
The electron-hole pair dissociates when electron or hole leaves the bottleneck. The probability of triplet exciton formation
$\gamma_t$ is neglected in comparison to $W_1^{(out)} + W_2^{(out)}$.

The probability of appearance of $t_0$ state is not affected by spin polarization of trapping sites leading to the
stationary condition for $p_0$
\begin{equation}
\frac{p_1 }{4}W_2^{(in)} + \frac{p_2}{4}
W_1^{(in)} = (W_1^{(out)} + W_2^{(out)})p_0
\end{equation}

The stationary conditions for the probabilities $p_1$ and $p_2$ lead to the equations
\begin{subequations}\label{6eq-12}
\begin{multline}
 W_2^{(out)}(p_+ + p_0) + W_1^{(in)}(1-p_1 - p_2 - p_+ - p_0) \\ = (W_1^{(out)} + W_2^{(in)})p_1
\end{multline}
\begin{multline}
 W_1^{(out)}(p_+ + p_0) + W_2^{(in)}(1-p_1 - p_2 - p_+ - p_0) \\ = (W_2^{(out)} + W_1^{(in)})p_2
\end{multline}
\end{subequations}

The stationary conditions for $p_{1s}$ and $p_{2s}$ read
\begin{subequations}\label{6eq-s}
\begin{equation}
W_2^{(out)}p_+ = (W_1^{(out)} + W_2^{(in)})p_{1s}
\end{equation}
\begin{equation}
W_1^{(out)}p_+ = (W_2^{(out)} + W_1^{(in)})p_{2s}
\end{equation}
\end{subequations}

The system of equations (\ref{6eq+}-\ref{6eq-s}) can be solved analytically but the solution is quite cumbersome. Here it is given
for the specific ``symmetrical'' case $W_1^{(in)} = W_2^{(in)} = W^{(in)}$, $W_1^{(out)} = W_2^{(out)} = W^{(out)}$.
In this case $p_1 = p_2$ and $p_{1s}=p_{s2}$. All the $p_\zeta$ are functions of the ratio $w=W^{(out)}/W^{(in)}$.

\begin{figure}[t]
  \centering
  \includegraphics[width=0.4\textwidth]{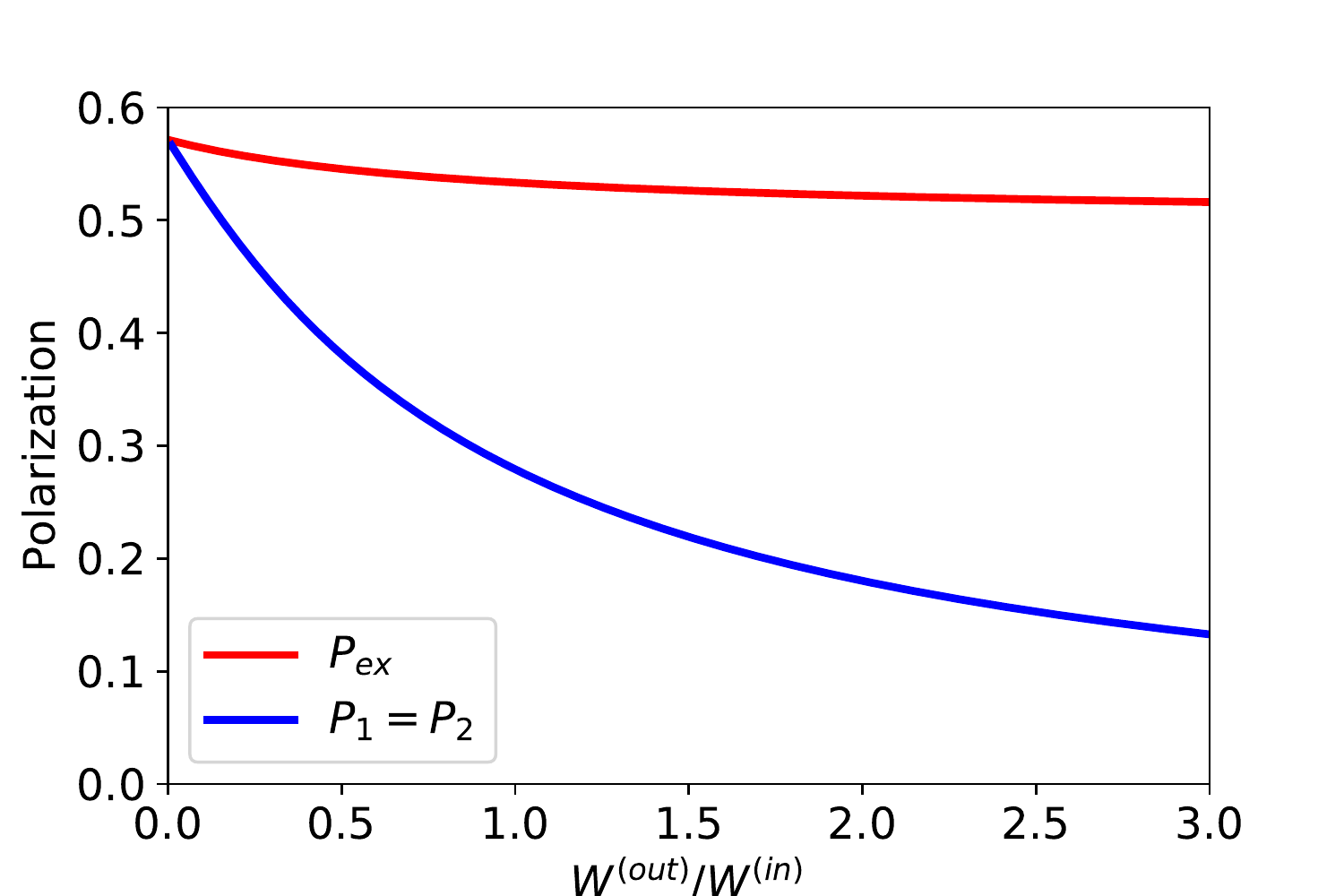}
  \caption{Spin polarizations $P_{ex}$, and $P_1 = P_2$ under the resonance conditions.}
  \label{fig:PolR}
\end{figure}

The spin polarization $p_{1s}$ of the site $1$ is related to the triplet occupation number $p_+$ as follows
\begin{equation}
p_{1s} = \frac{w p_+}{w+1}
\end{equation}
The triplet occupation probabilities are proportional to the probability $p_1$ of the occupation of a single site.
\begin{equation}
p_0 = \frac{p_1}{4w}, \quad p_+=\frac{p_1}{4w} \frac{1+w}{3/4+w}
\end{equation}
The analytical expression for $p_1$ is
\begin{equation}
p_1 = \frac{4w(4w+3)}{16w^3+52w^2+37w+7}
\end{equation}

The spin polarizations $P_1$, $P_2$ and $P_{ex}$ than should be calculated from Eqs.~(\ref{Pol-p}). The result of such a calculation is shown in
Fig.~\ref{fig:PolR}. When $W^{(in)} \gg W^{(out)}$ the polarizations $P_1 = P_2$ and $P_{ex}$ coincide.
 Both of the trapping sites are almost always occupied and $p_1$ can be neglected
in comparison to $p_0$ and $p_+$. $p_+$ is equal to $4p_0/3$ in this case leading to $P_{ex} = P_1=P_2 = 4/7$. It is the
largest spin polarization possible in the discused model. Note that in Sec.~\ref{sec:pol} $W_{1}^{(in)}\gg W_2^{(out)}$
and $W_{2}^{(in)} \gg W_{2}^{(out)}$ were considered. Fig.~\ref{fig:Pol1} shows that polarizations $P_1$, $P_2$ and $P_{ex}$
coincide in this case even for non-resonance situation.

In the opposite limit $W^{(out)} \gg W^{(in)}$ the probabilities of $t_+$ and $t_0$ states are equal leading to
$P_{ex} = 1/2$. Polarizations $P_1$ and $P_2$ are small in this limit.

\section{discussion}
\label{sec:dis}

The dynamic spin polarization was observed in non-organic GaAs quantum dots due to the circular polarization of
photoluminescence. It was possible due the strong spin-orbit intercation in GaAs that allows radiative recombination
of excitons with angular momentum equal to unity. In organic semiconductors such a detection is not an easy task because usually only singlet
excitons recombine radiatively. The radiative recombination of triplets in organic is called phosphorescence and
can be achieved by the introduction of certain impurities (the so-called phosphors) to organic semiconductors.
It makes the optical detection of dynamic spin polarization  in organics possible at least in theory. However, such a detection is related
to additional restrictions that are out of the scope of the provided model: the spin of a triplet exciton should be
conserved during the transition to such a phosphor and the following recombination process.

However, organic semiconductors also have advantages over the quantum dots when dynamic spin polarization is
considered. Spin-phonon interaction in non-organic semiconductors leads to fast spin relaxation that suppress circular
polarization of luminescence in quantum dots at temperatures $T\gtrsim 10\,{\rm K}$ \cite{Smirnov125}.
However, OMAR and strong spin-valve effect exist both at low and room temperatures due to the weak spin-orbit interaction
in organic materials.
It makes it possible for dynamic spin polarization to also exist at room temperature.

The long spin diffusion length measured in some organic semiconductors gives hope that the spin polarization can be detected
 in transport
measurements in hybrid devices with ferromagnetic contacts. Another possibility is the muon spin rotation experiments
that were able to detect spin-polarization in working spin-valve devices. Actually the author believes that
it may be relevant to revisit organic spin-valve experiments in view of the results of this article. Usually only
the two possibilities were considered for organic spin-valve: the spin is injected from the first ferromagnetic contact and
is detected by the second one or the spin-valve is due to the tunneling through pin-holes in organic layer.
Now the third assumption should be
added: the spin can be generated inside organic layer. The external fields $\sim 100\,{\rm mT}$ required for such a
generation can be related for example to fringe fields of magnetic contacts \cite{Fringe}.

The effective dynamic spin polarization requires exchange energies between electrons and holes on different molecules
in the bottleneck to be larger or comparable with the energy of hyperfine interaction of electron and nuclear spins.
The existing estimates of the exchange energy are quite controversial. In \cite{IsotopePRB} the value $B_{ex} = 0.2\,{\rm mT}$
was extracted from the comparison of measured OMAR shape with theory. Such a value is clearly insufficient for the
significant spin polarization. However such an estimate of the exchange energy can be complicated if $B_{ex}$
has broad distribution, as it follows from Fig.~\ref{fig:OMAR-distr}. The exchange interaction between electrons
localized on different molecules was also invoked in Ref.~\cite{Yu} to describe the absence of Hanle effect in organic
spin-valves. The exchange energies corresponding to $B_{ex}\gtrsim 0.1T$ where considered to be possible. Perhaps the
typical exchange energies in organic semiconductor can be very different in different samples and depend not only on
chemical structure but also on the concentration of injected charge carriers. The strong dynamic spin polarization can occur in
the samples where the condition $B\sim B_{ex} \gtrsim B_{hf}$ is satisfied.

In conclusion it was shown that exchange interaction between electrons and holes localized on different molecules
leads to spin polarization in organic semiconductors where OMAR is observed. For the polarization to be significant
the exchange interaction should be comparable or larger than hyperfine interaction of electron and nuclear spins.
The exchange interaction also modifies the shape of OMAR. However, such a modification can be masked by a broad
distribution of exchange energies. This broad distribution does not completely suppress the spin polarization.

The author is grateful to D.S. Smirnov, V.V. Kabanov, V.I. Dediu and Y.M. Beltukov for many fruitful discussions.
The support from Foundation for the Advancement of Theoretical Physics and Mathematics ``Basis'' is greatly
acknowledged.

\appendix

\section{Magnetoresistance in electron-hole mechanism}
\label{appen-eh}

To calculate the magnetoresistance in electron-hole mechanism the current should be expressed as the rate of trapping
in sites $1$ and $2$
\begin{equation}\label{eh-Jc1}
J = e(1 - \overline{n}_1) W_{1}^{(in)} - e W_1^{(out)} \overline{n_1}
\end{equation}
\begin{equation} \label{eh-Jc2}
J = e(1 - \overline{n}_2) W_{2}^{(in)} - e W_2^{(out)} \overline{n_2}
\end{equation}
Eq.~(\ref{eh-Jc1}) shows that the electron cannot be trapped on site $1$ if it is already occupied
and that all the electrons that are trapped and do not leave site $1$ contribute to exciton generation
and current. Eq.~(\ref{eh-Jc2}) is the similar expression for site $2$.

The current given by Eqs.~(\ref{eh-Jc1},\ref{eh-Jc2}) is equal to the current (\ref{eh-J1}). This system of
equation should be complemented by the master equation for joined occupation probability $\overline{n_1 n_2}$
\begin{multline}\label{eh-n1n2}
\frac{d}{dt} \overline{n_1 n_2} = 0 =
W_{1}^{(in)} \overline{n}_2 + W_{2}^{(in)} \overline{n}_1 \\  - \left[\widetilde{W}_{12} + \gamma_{eh}(B)\right] \overline{n_1 n_2}
\end{multline}
Here $\widetilde{W}_{12} = W_1^{(in)} + W_1^{(out)} + W_2^{(in)} + W_2^{(out)}$.

The solution of Eqs.~(\ref{eh-J1}, \ref{eh-Jc1} - \ref{eh-n1n2}) yields the following expressions for
$n_1$ and $n_2$:
\begin{equation}
\overline{n}_1 = \frac{W_1^{(in)} - \widetilde{\gamma}\overline{n_1n_2}}{W_1^C}
\end{equation}
\begin{equation}
\overline{n}_2 = \frac{W_2^{(in)} - \widetilde{\gamma}\overline{n_1n_2}}{W_2^C}
\end{equation}
Here $W_1^C = W_1^{(in)} + W_1^{(out)}$, $W_2^C = W_2^{(in)} + W_2^{(out)}$.

The current $J$ and joined occupation probability $\overline{n_1n_2}$ are expressed as follows:
\begin{equation}\label{appen-J}
J = \frac{e\gamma_{eh} W_1^{(in)}W_2^{(in)}\widetilde{W}_{12}}{W_1^CW_2^C \widetilde{W}_{12}+\gamma_{eh}{\cal W}}
\end{equation}
\begin{equation}
\overline{n_1n_2} = \frac{W_1^{(in)}W_2^{(in)}\widetilde{W}_{12}}{W_1^CW_2^C \widetilde{W}_{12}+\gamma_{eh}{\cal W}}
\end{equation}
Here ${\cal W} = W_1^C W_2^C + W_1^C W_1^{(in)} + W_2^C W_2^{(in)}$.

The magnetoresistance can be expressed as $(J(0) - J(B))/J(B)$. When the effect of spin correlations
on conductivity is small, the magnetoresistance is also small
($|J(0) - J(B)| \ll J(0)$) and  Eq.~(\ref{MR-gen}) can be derived from Eq.~(\ref{appen-J}) with $C_{eh}$ equal to
\begin{equation}\label{eh-C}
C_{eh} = \frac{W_1^{C}W_2^{C}\widetilde{W}_{12}}{W_1^CW_2^C \widetilde{W}_{12}+\gamma_{eh}(0){\cal W}}
\end{equation}
Here $\gamma_{eh}(0)$ is calculated in zero magnetic field.

\end{document}